\documentclass[12pt]{article}
 
\def\datetitle{September 2, 1998}
\def\datehead{September 2, 1998}
 
\makeatletter
 
\@addtoreset{equation}{section}

\begin{document}

    \def\@evenhead{\rm\small\thepage\hfil\datehead}
    \def\@oddhead{\datehead\hfil\rm\small\thepage}

%
%
 
\makeatother
 
%
%
 
\def\bbbr{{\rm I\!R}} 
\def\bbbn{{\rm I\!N}} 
\def\bbbm{{\rm I\!M}}
\def\bbbh{{\rm I\!H}}
\def\bbbf{{\rm I\!F}}
\def\bbbk{{\rm I\!K}}
\def\bbbp{{\rm I\!P}}
\def\bbbone{{\mathchoice {\rm 1\mskip-4mu l} {\rm 1\mskip-4mu l}
{\rm 1\mskip-4.5mu l} {\rm 1\mskip-5mu l}}}
\def\bbbc{{\mathchoice {\setbox0=\hbox{$\displaystyle\rm C$}\hbox{\hbox
to0pt{\kern0.4\wd0\vrule height0.9\ht0\hss}\box0}}
{\setbox0=\hbox{$\textstyle\rm C$}\hbox{\hbox
to0pt{\kern0.4\wd0\vrule height0.9\ht0\hss}\box0}}
{\setbox0=\hbox{$\scriptstyle\rm C$}\hbox{\hbox
to0pt{\kern0.4\wd0\vrule height0.9\ht0\hss}\box0}}
{\setbox0=\hbox{$\scriptscriptstyle\rm C$}\hbox{\hbox
to0pt{\kern0.4\wd0\vrule height0.9\ht0\hss}\box0}}}}
\def\bbbe{{\mathchoice {\setbox0=\hbox{\smalletextfont e}\hbox{\raise
0.1\ht0\hbox to0pt{\kern0.4\wd0\vrule width0.3pt
height0.7\ht0\hss}\box0}}
{\setbox0=\hbox{\smalletextfont e}\hbox{\raise
0.1\ht0\hbox to0pt{\kern0.4\wd0\vrule width0.3pt
height0.7\ht0\hss}\box0}}
{\setbox0=\hbox{\smallescriptfont e}\hbox{\raise
0.1\ht0\hbox to0pt{\kern0.5\wd0\vrule width0.2pt
height0.7\ht0\hss}\box0}}
{\setbox0=\hbox{\smallescriptscriptfont e}\hbox{\raise
0.1\ht0\hbox to0pt{\kern0.4\wd0\vrule width0.2pt
height0.7\ht0\hss}\box0}}}}
\def\bbbq{{\mathchoice {\setbox0=\hbox{$\displaystyle\rm Q$}\hbox{\raise
0.15\ht0\hbox to0pt{\kern0.4\wd0\vrule height0.8\ht0\hss}\box0}}
{\setbox0=\hbox{$\textstyle\rm Q$}\hbox{\raise
0.15\ht0\hbox to0pt{\kern0.4\wd0\vrule height0.8\ht0\hss}\box0}}
{\setbox0=\hbox{$\scriptstyle\rm Q$}\hbox{\raise
0.15\ht0\hbox to0pt{\kern0.4\wd0\vrule height0.7\ht0\hss}\box0}}
{\setbox0=\hbox{$\scriptscriptstyle\rm Q$}\hbox{\raise
0.15\ht0\hbox to0pt{\kern0.4\wd0\vrule height0.7\ht0\hss}\box0}}}}
\def\bbbt{{\mathchoice {\setbox0=\hbox{$\displaystyle\rm
T$}\hbox{\hbox to0pt{\kern0.3\wd0\vrule height0.9\ht0\hss}\box0}}
{\setbox0=\hbox{$\textstyle\rm T$}\hbox{\hbox
to0pt{\kern0.3\wd0\vrule height0.9\ht0\hss}\box0}}
{\setbox0=\hbox{$\scriptstyle\rm T$}\hbox{\hbox
to0pt{\kern0.3\wd0\vrule height0.9\ht0\hss}\box0}}
{\setbox0=\hbox{$\scriptscriptstyle\rm T$}\hbox{\hbox
to0pt{\kern0.3\wd0\vrule height0.9\ht0\hss}\box0}}}}
\def\bbbs{{\mathchoice
{\setbox0=\hbox{$\displaystyle     \rm S$}\hbox{\raise0.5\ht0\hbox
to0pt{\kern0.35\wd0\vrule height0.45\ht0\hss}\hbox
to0pt{\kern0.55\wd0\vrule height0.5\ht0\hss}\box0}}
{\setbox0=\hbox{$\textstyle        \rm S$}\hbox{\raise0.5\ht0\hbox
to0pt{\kern0.35\wd0\vrule height0.45\ht0\hss}\hbox
to0pt{\kern0.55\wd0\vrule height0.5\ht0\hss}\box0}}
{\setbox0=\hbox{$\scriptstyle      \rm S$}\hbox{\raise0.5\ht0\hbox
to0pt{\kern0.35\wd0\vrule height0.45\ht0\hss}\raise0.05\ht0\hbox
to0pt{\kern0.5\wd0\vrule height0.45\ht0\hss}\box0}}
{\setbox0=\hbox{$\scriptscriptstyle\rm S$}\hbox{\raise0.5\ht0\hbox
to0pt{\kern0.4\wd0\vrule height0.45\ht0\hss}\raise0.05\ht0\hbox
to0pt{\kern0.55\wd0\vrule height0.45\ht0\hss}\box0}}}}
 
%
%
 
\def\bbbz{{\mathchoice {\hbox{$\sf\textstyle Z\kern-0.4em Z$}}
{\hbox{$\sf\textstyle Z\kern-0.4em Z$}}
{\hbox{$\sf\scriptstyle Z\kern-0.3em Z$}}
{\hbox{$\sf\scriptscriptstyle Z\kern-0.2em Z$}}}}
 
%
%
\newtheorem{theorem}{Theorem}[section]          
\newtheorem{lemma}[theorem]{Lemma}              
\newtheorem{proposition}[theorem]{Proposition}
\newtheorem{corollary}[theorem]{Corollary}
\newtheorem{definition}[theorem]{Definition}
\newtheorem{conjecture}[theorem]{Conjecture}
\newtheorem{claim}[theorem]{Claim}
\newtheorem{observation}[theorem]{Observation}
\def\proof{\par\noindent{\it Proof.\ }}
\def\reff#1{(\ref{#1})}
%
%
\let\zed=\bbbz 
\let\szed=\bbbz 
\let\IR=\bbbr 
\let\R=\bbbr 
\let\sIR=\bbbr 
\let\IN=\bbbn 
\let\IC=\bbbc 

\def\nl{\medskip\par\noindent}
 
\def\scrb{{\cal B}}
\def\scrg{{\cal G}}
\def\scrf{{\cal F}}
\def\scrl{{\cal L}}
\def\scrr{{\cal R}}
\def\scrt{{\cal T}}
\def\pfin{{\cal S}}
\def\prob{M_{+1}}
\def\cql{C_{\rm ql}}
\def\bydef{:=}
\def\qed{\hbox{\hskip 1cm\vrule width6pt height7pt depth1pt \hskip1pt}\bigskip}
\def\remark{\medskip\par\noindent{\bf Remark:}}
\def\remarks{\medskip\par\noindent{\bf Remarks:}}
\def\example{\medskip\par\noindent{\bf Example:}}
\def\examples{\medskip\par\noindent{\bf Examples:}}
\def\nonexamples{\medskip\par\noindent{\bf Non-examples:}}

\newenvironment{scarray}{
          \textfont0=\scriptfont0
          \scriptfont0=\scriptscriptfont0
          \textfont1=\scriptfont1
          \scriptfont1=\scriptscriptfont1
          \textfont2=\scriptfont2
          \scriptfont2=\scriptscriptfont2
          \textfont3=\scriptfont3
          \scriptfont3=\scriptscriptfont3
        \renewcommand{\arraystretch}{0.7}
        \begin{array}{c}}{\end{array}}
 
\def\wspec{w'_{\rm special}}
\def\mup{\widehat\mu^+}
\def\mupm{\widehat\mu^{+|-_\Lambda}}
\def\pip{\widehat\pi^+}
\def\pipm{\widehat\pi^{+|-_\Lambda}}
\def\ind{{\rm I}}
\def\const{{\rm const}}
\def\ft#1{\widehat{#1}} 
\def\vv#1{\vec{#1}} 
\def\tends#1{\mathop{\longrightarrow}\limits_{#1}}
\def\ssum{\mathop{\hbox{``}\sum\nolimits\hbox{''}}\limits}
\def\hyp#1#2{\smallskip\par\noindent{\bf (#1)  #2}.}

\bibliographystyle{plain}
 
 
\title{\vspace*{-2.4cm} Problems with the definition\break
of renormalized Hamiltonians\break
for momentum-space renormalization transformations}

\author{
  \\
  {\normalsize Aernout C. D. van Enter}        \\[-1.5mm]
  {\normalsize\it Institute for Theoretical Physics}   \\[-1.5mm]
  {\normalsize\it Rijksuniversiteit Groningen}         \\[-1.5mm]
  {\normalsize\it Nijenborgh 4}                \\[-1.5mm]
  {\normalsize\it 9747 AG Groningen}           \\[-1.5mm]
  {\normalsize\it THE NETHERLANDS}             \\[-1mm]
  {\normalsize\tt AENTER@PHYS.RUG.NL}        \\[-1mm]
  {\protect\makebox[5in]{\quad}}  
  \\[-1mm] \and {\normalsize Roberto Fern\'andez\thanks{Researcher of
      the National Research Council [Consejo Nacional de
      Investigaciones Cient\'{\i}ficas y T\'ecnicas (CONICET)],
      Argentina.}
    }\\
{\normalsize\it Instituto de Estudos Avan\c{c}ados, }\\[-1.5mm]
{\normalsize\it Universidade de S\~ao Paulo}\\[-1.5mm]
{\normalsize\it Av.\ Prof.\ Luciano Gualberto, }\\[-1.5mm]
{\normalsize\it Travessa J, 374 T\'erreo} \\[-1.5mm]
{\normalsize\it 05508-900 - S\~{a}o Paulo, Brazil}\\[-1mm]
{\normalsize\tt rf@ime.usp.br}\\[-2mm]
}
 
\date{\datetitle}
 
\maketitle
\thispagestyle{empty}

\clearpage

\setcounter{page}{1}
 
\begin{abstract}
For classical lattice systems with finite (Ising) spins,
we show that the implementation of momentum-space renormalization at
the level of Hamiltonians runs into the same type of difficulties as
found for real-space transformations:  Renormalized Hamiltonians are
ill-defined in certain regions of the phase diagram.  
\end{abstract}
\medskip



\section{Introduction}
Despite the great success of renormalization-group (RG) ideas, both for
computations and as a heuristic guide, 
many aspects of the theory still lack rigorous mathematical
justification.  The filling of this gap is more than just of
academic interest.  It has been repeatedly pointed out 
(e.g.\ \cite[p.~82]{fis83}, \cite[footnote in page 38]{bengal95},
\cite[p.~268]{gol92}) that the method is not a black-box type of
technique; its succesful application requires some understanding
of the underlying physics or one may be led to incorrect conclusions.
Studies on the foundations of real-space transformations 
 \cite{gripea78,gripea79,isr79,vEFS_PRL,vEFS_JSP} suggest that a
similar remark applies to the underlying mathematics.  Indeed,
these studies show that in various occasions renormalized Hamiltonians
are ill-defined.  The finite-volume probabilities of the renormalized
system exhibit a long-range dependence on boundary
spins that is incompatible with the existence of a Hamiltonian, at
least one defined in the usual (summable) sense.  Such a ``pathology''
is usually referred to as {\em non-Gibbsianness}\/.  This
phenomenon, which appears after a {\em single}\/ application of the
transformation, was first detected in the vicinity
of first-order phase transitions, but was later discoverd in other
regions of phase diagrams, including at high magnetic fields
\cite{vEFS_JSP,vEFK} and at high temperatures \cite{vEFK,entpre}.  
It follows that the design of the renormalization transformation is
crucial for the very {\em existence}\/ of a renormalization flow in a
suitable space.

Nevertheless, the lack of similar studies for momentum-space
transformations left open the possibility that they could be free of
this pathological behavior.  That is, the question remained as to
whether such transformations, possibly with a soft cutoff, would
generally lead to a honest-to-God renormalized Hamiltonian
\cite{gripea79,fis92}.  In this note we present a simple example
showing that this is in general {\em not}\/ the case, as already
suspected by Griffiths \cite{gri81}.  There is no essential difference
between real-space and momentum-space transformations at least in case
the spins are bounded.  The same mechanism ---the existence of a phase
transition in the system of original spins, constrained by a
particular block-spin configuration--- causes similar problems with
the definition of renormalized Hamiltonians.  For an earlier
suggestion that momentum-space maps are not all that different from
real-space maps, see also \cite{hashas88}.

As we remark at the end of Section \ref{s.non}, these problems
can be interpreted as a manifestation of the well-known ``large-field
problem''.  It might be hoped that the considerable experience
accumulated in the treatment of this type of problems, could be of
help to control the non-Gibbsianness ``pathologies''.

\section{Momentum transformations}\label{s.mom}

We consider Ising spins $\sigma_{\vv x}=-1,+1$ on a lattice, $x\in\zed^d$.
For each finite periodic cube $V=[-N,N]^d$ in $\zed^d$ we define the
Fourier-transformed variables  
\begin{equation}
\ft{\sigma}^V_{\vv k} \;\bydef\; 
\sum_{\vv x\in V} \sigma_{\vv x} \,
e^{-i \vv k \vv x}\;,
\label{ft.1}
\end{equation}
where $\vv k \vv x\bydef k_1x_1+\cdots+k_dx_d$, and each $k_i$ belongs
to the Brillouin zone: 
${\cal B}_N=\{-\pi, -\pi[1-1/(2N+1)], \ldots, \pi[1-1/(2N+1)], \pi\}$.
The inverse of \reff{ft.1} is
\begin{equation}
{\sigma}_{\vv x} \;=\; {1\over (2N+1)^d} 
\sum_{\vv k\in {\cal B}_N^d} \ft\sigma^V_{\vv k} \,
e^{i \vv k \vv x}\;,
\label{ft.1.1}
\end{equation}
for $\vv x\in V$.  

A {\em momentum-space transformation}\/ is defined in two steps:
\begin{itemize}
\item[{\em 1st}\/.]  A cutoff is applied to the variables
$\ft\sigma^V_{\vv k}$:
\begin{equation}
\ft\sigma^{\prime V}_{\vv k} \;\bydef\; \ft f(\vv k) \, \ft\sigma^V_{\vv k}\;.
\label{ft.5}
\end{equation}
The {\em volume-independent}\/ cutoff function $\ft f$ is designed so
as to keep only momenta smaller than a certain threshold $k_0$.  
\item[{\em 2nd}\/.]  Momenta are rescaled by the factor $k_0$ so
as to return to a Brillouin zone in $[-\pi,\pi]^d$:
\begin{equation}
\ft\sigma^{\prime V}_{\vv k'} \;\bydef\; \ft f(\vv k'k_0/\pi) \,
\ft\sigma^V_{\vv k'k_0/\pi}\;.
\label{ft.10}
\end{equation}
\end{itemize}
In addition, the renormalized variables $\ft\sigma^{\prime V}_{\vv k'}$
are usually rescaled in applications.  We will not do this, as we
shall not apply the transformation more than once.  

In Wilson's original approach 
\cite[and references therein]{Wilson-Kogut}, 
the cutoff function $\ft f(\vv k)$ was chosen
simply as the step function
\begin{equation}
\chi_{k_0}(\vv k) \;\bydef\; \left\{\begin{array}{ll}
1 & \hbox{if } |k_i|\le k_0 \; i=1,\ldots, d\\
0 & \hbox{otherwise}\;.
                                \end{array}
\right.
\label{ft.15}
\end{equation}
It was quickly realized, however, that such a sharp cutoff leads to
unwanted long-range terms in the renormalized Hamiltonian
(see e.g.\ \cite[p.\ 153]{Wilson-Kogut}, \cite[Appendix 2]{ma76}).
To avoid such terms one usually takes smooth momentum-cutoffs, that
is, functions $\ft f$ which go to zero in a sufficiently
differentiable fashion.  Such functions are obtained, for instance, via
a convolution
\begin{equation}
\ft f(\vv k) \;=\; \sum_{\vv\ell\in{\cal B}_N^d} 
\delta_\Delta(\vv k-\vv\ell)\, \chi_{k_0}(\vv k)
\label{ft.16}
\end{equation}
with a smooth delta-like function $\delta_\Delta$ peaked
at $k=0$ of width $\Delta$.  

Rigorously speaking, one is interested in the limit $V\to\zed^d$ of
this procedure.  To make sense of this limit we return to 
real space, where the prescription \reff{ft.10}
translates into the relation
\begin{equation}
\sigma^{\prime V}_{\vv x'} \;=\; \sum_{\vv y\in V} f^V(L\vv x'-\vv y)\, 
\sigma_{\vv y}\quad,\quad x'\in V/L\;,
\label{ft.20}
\end{equation}
where $f^V$ is the $V$-dependent (inverse) discrete Fourier transform
of $\ft f$ and
\begin{equation}
L \;\bydef\; {\pi\over k_0}\;.
\label{ft.21}
\end{equation}
The volume $V$ is assumed to be a disjoint union of cubes of side $L$
(i.e.\ $N$ is a multiple of $L$).  As $V\to\zed^d$, the function
$f^V(\vv x)$ tends to
\begin{equation}
f(\vv x) \;\bydef\; {1\over (2\pi)^d} \int_{-\pi}^\pi \ft f(\vv k) \, 
e^{-i \vv k \vv x}\, d\vv k\;.
\label{ft.ft}
\end{equation}
A sharp cutoff in momentum space gives rise to a {\em non-summable}\/
function $f$, i.e.\ $\sum_{\vv x\in\zed^d} |f(\vv x)|=\infty$.
[The inverse transform of $\chi_{k_0}$ is
proportional to the function $\prod_{i=1}^d\sin(k_0 x_i)/(k_0 x_i)$].
Summability is 
restored if $\ft f$ is smooth enough (for example once differentiable).  
In such a case, expression
\reff{ft.20} remains valid in the thermodynamic limit:
\begin{equation}
\sigma'_{\vv x'} \;=\; \sum_{\vv y\in \zed^d} f(L\vv x'-\vv y)\, 
\sigma_{\vv y}\;,
\label{ft.20.1}
\end{equation}
and
the renormalized spins remain {\em bounded}\/ in this limit.
They may take a large number of values, but within some finite interval.

Expression \reff{ft.20.1} shows that a cutoff in momentum space, even
a smooth one like \reff{ft.16}, leads to {\em non-local}\/ averages in
real space, i.e.\ to functions $f$ extending to infinity.  This is the
distinctive feature with respect to the real-space transformations
analyzed, for instance, in \cite{vEFS_JSP}.  Nevertheless, it is
expected that ``the physics behind integration over fluctuations
having wavenumbers [$|k_i|>k_0$] is the same as the physics behind the
formation of blocks of spins having volume [$L^d$] in real space''
\cite[Section 4.2]{pfetou77}.  To ensure this, the momentum-space
cutoff should lead to an {\em almost local}\/ average in real space.
That is, the function $f$ should decay rapidly outside of a region of
size not much larger than $L$.  We see that if $f$ has a Fourier
transform of the type \reff{ft.16}, the contribution of spins outside
the volume of size $L$ is of order $\ln (\Delta/k_0)$.  We conclude
that the cutoff function $\ft f$ must approach zero in a ``gradual''
manner, that is, with $\Delta$ of the order of $k_0$ in \reff{ft.16}.

A {\em soft momentum-cutoff}\/ is a function $\ft f$ that is smooth
and gradual in the above sense.

\section{The phenomenon of non-Gibbsianness}

A state (probability measure or distribution) is called Gibbsian
if it can be written in terms of Boltzmann-Gibbs weights for an
``acceptable Hamiltonian [which] \ldots must satisfy the additional
requirement of locality \ldots [that is,] a quantity that is additive
over distant lattice sites'' \cite[p.\ 145]{Wilson-Kogut}.  In other
words, Hamiltonians must be such that the energy of disjoint
volumes is additive except for boundary terms whose contribution is 
small in comparison with the volumes.  For classical lattice systems,
the appropriate requirement is that the flipping of one spin lead to a
{\em finite}\/ energy change whatever the configuration of the
remaining spins is.  This intuition has been precisely formalized in
the form of a summability condition that we now present.

Hamiltonians are, in general, sums of many-body terms:
\begin{equation}
H(\sigma) \;=\; \sum_B \Phi_B(\sigma_B)\;,
\label{ft.40}
\end{equation}
where each $\Phi_B$ is a function only of the spins in the finite set
$B\subset\zed^d$, i.e.\ of the variables 
$\sigma_B=\{\sigma_{\vv x}\}_{\vv x\in B}$.  For Ising spins, these functions
$\Phi_B$ are usually written in the form $J_B\sigma_B$; the general
expression \reff{ft.40} is more suitable for spins
larger than 1/2, where one would need powers of $\sigma_{\vv x}$,
and also for some particular spin-1/2 interactions \cite{vanfer89}.
The finiteness of the single-flip energy change translates into the
following summability requirement:
\begin{equation}
\sup_{\vv x} \sum_{B\ni \vv x} \|\Phi_B\| \;<\; \infty\;,
\label{ft.sum}
\end{equation}
where $\|\Phi_B\|\bydef \sup_\sigma |\Phi_B(\sigma_B)|$.

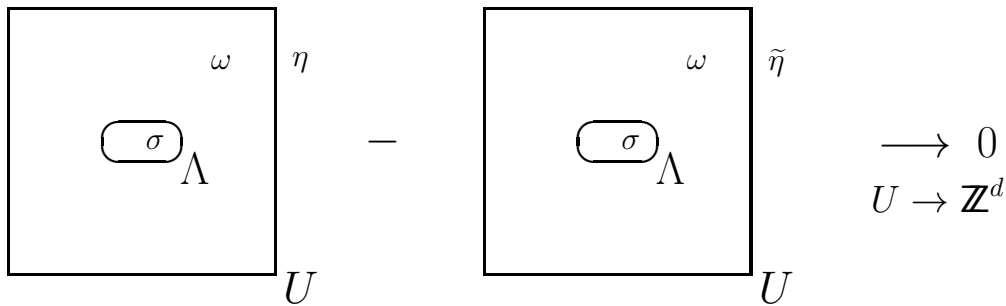
\begin{figure}
\begin{center}
\begin{picture}(340,110)(-140,-55)
\thicklines
%
\put(-165,-50){
\begin{picture}(110,110)(-50,-50)
\thicklines
\put(0,0){\oval(30,15)}
\put(5,0){\makebox(0,0){$\sigma$}}
\put(20,-10){\makebox(0,0){\Large$\Lambda$}}
\put(30,30){\makebox(0,0){$\omega$}}
\put(-50,-50){\framebox(100,100){\ }}
\put(60,-55){\makebox(0,0){\Large$U$}}
\put(60,30){\makebox(0,0){$\eta$}}
\end{picture}}
\put(-20,0){\makebox(0,0){\Large$-$}}
\put(15,-50){
\begin{picture}(110,110)(-50,-50)
\thicklines
\put(0,0){\oval(30,15)}
\put(5,0){\makebox(0,0){$\sigma$}}
\put(20,-10){\makebox(0,0){\Large$\Lambda$}}
\put(30,30){\makebox(0,0){$\omega$}}
\put(-50,-50){\framebox(100,100){\ }}
\put(60,-55){\makebox(0,0){\Large$U$}}
\put(60,30){\makebox(0,0){$\widetilde\eta$}}
\end{picture}}
%
%
\put(190,0){\makebox(0,0){\Large$\longrightarrow\; 0$}}
\put(190,-20){\makebox(0,0){\large$U \to \zed^d$}}
\end{picture}
\end{center}
\caption{Test for quasilocality.  As $U$ tends to $\zed^d$ while keeping
fixed the intermediate configuration 
$\omega$, the energy inside $\Lambda$ should asymptotically become
independent of the configuration $\eta$ or $\widetilde\eta$ outside $U$}
\label{fig.1}
\end{figure}

If this summability holds, then we can consider Hamiltonians with
arbitrary boundary conditions.  For each cube $\Lambda$ in
$\zed^d$, let
\begin{equation}
H_\Lambda(\sigma|\omega) \;=\; \sum_{B: B\cap\Lambda\neq\emptyset}
\Phi_B\bigl((\sigma_\Lambda\omega)_B\bigr)\;,
\label{ft.45}
\end{equation}
where $\sigma_\Lambda\omega$ is the configuration
\begin{equation}
(\sigma_\Lambda\omega)_{\vv x} \;=\; \left\{\begin{array}{ll}
\sigma_{\vv x} & \hbox{if } \vv x \in \Lambda\\
\omega_{\vv x} & \hbox{if } \vv x \in \zed^d\setminus\Lambda\;.
                                            \end{array}
\right.
\label{ft.50}
\end{equation}
For finite-range interactions, the dependence of the Hamiltonian
\reff{ft.45} on the boundary condition $\omega$ is only through spins
at sites not farther from $\Lambda$ than the range of the interaction.
More generally, even if the Hamiltonian involves terms with
arbitrarily long range, the summability condition \reff{ft.sum}
implies that the dependence on the boundary spins decays with their
distance to the volume 
$\Lambda$.  This property, which is called {\em quasilocality}\/,
is central for our argument.  Lets us state it precisely.  We
take a sequence of cubes $U\supset\Lambda$ with larger and larger
radius and {\em fix}\/ a configuration $\omega$ in the intermediate
(``buffer'') region $U\setminus\Lambda$ (see Figure \ref{fig.1}).  It
is not hard to see that the summability condition implies (in fact, it
is equivalent to) the following fact:
As $U$ tends to the whole $\zed^d$ the Hamiltonian becomes independent
of what happens outside $U$:
\begin{equation}
\sup_{\eta\,\widehat\eta} \Bigl|H_\Lambda(\sigma|\omega_U\eta) -
H_\Lambda(\sigma|\omega_U\widehat\eta)\Bigr| \;\tends{U\to\zed^d}\; 0
\label{ft.55}
\end{equation}
for all $\sigma_\Lambda$ and all boundary conditions $\omega$, for all
cubes $\Lambda$.  Hamiltonians $H_\Lambda$ satisfying \reff{ft.55} are
said \emph{quasilocal}.

The condition \reff{ft.55} is appropriate for discrete finite spins.
The renormalized spins considered in the next section, however, can
take an infinite number of values.  In such a situation the
quasilocality condition must allow small variations in the spin
configurations $\sigma$ and $\omega$ \cite{nondiscrete}.

The Hamiltonians \reff{ft.45} are used to construct the Boltzmann-Gibbs
weights 
\begin{equation}
\rho_\Lambda(\sigma|\omega) \;\bydef\;
{\exp[-\beta H_\Lambda(\sigma|\omega)] \over Z_\Lambda(\omega)}
\label{ft.60}
\end{equation}
[$Z_\Lambda(\omega)$ is the obvious normalization factor],
which do inherit the quasilocality property \reff{ft.55}:
\begin{equation}
\sup_{\eta\,\widetilde\eta} \Bigl|\rho_\Lambda(\sigma|\omega_U\eta) -
\rho_\Lambda(\sigma|\omega_U\widetilde\eta)\Bigr| 
\;\tends{U\to\zed^d}\; 0 \;,
\label{ft.65}
\end{equation}
for all cubes $\Lambda$, configurations $\sigma_\Lambda$ and boundary
conditions $\omega$.  For spins taking infinitely many values, the
condition should be slightly modified as described in note
\cite{nondiscrete}. 

A momentum-space renormalization amounts to passing to {\em
renormalized Boltzmann-Gibbs weights}\/
\begin{equation}
\rho'_{\Lambda'}(\sigma'|\omega') \;=\;
{1 \over Z_{\Lambda'}(\omega')}\,
\ssum_{\sigma_\Lambda\omega\, \in\, {\cal
C}(\sigma'_{\Lambda'}\omega')}
\exp[-\beta H_\Lambda(\sigma|\omega)] \;,
\label{ft.70}
\end{equation}
where ${\cal C}(\sigma'_{\Lambda'}\omega')$ denotes the set of
(original) spin configurations compatible with the ``block-spin''
configuration $\sigma'_{\Lambda'}\omega'$, that is the set of 
$\eta$ such that
\begin{equation}
\sum_{\vv y\in \zed^d} f(L\vv x'-\vv y)\, 
\eta_{\vv y} \;=\; \left\{\begin{array}{ll}
\sigma'_{\vv x'} & \hbox{if } \vv x' \in \Lambda'\\
\omega'_{\vv x'} & \hbox{if } \vv x' \in \zed^d\setminus\Lambda'\;.
                          \end{array}
\right.
\label{ft.75}
\end{equation}
The symbol $\ssum$ in \reff{ft.70} reminds us that the operation
involved is not a standard sum because we are ``summing'' over
uncountably many configurations.  The operation is, therefore, a sum
combined with a suitable limit procedure or, in mathematical terms, an 
integral with respect to the countable product measure 
$\prod_{\vv x\in \zed^d} [(1/2)\sum_{\eta_{\vv x}}]$.
(The reader interested in the rigorous construction of \reff{ft.70}
is referred to the discussion in \cite[pp.\ 987--90]{vEFS_JSP}.)

Most of the studies based on renormalization ideas assume that 
renormalized weights \reff{ft.75} can also be written as 
Boltzmann-Gibbs weights for an {\em acceptable}\/ ---in Wilson's and
Kogut's sense--- renormalized Hamiltonian.  That
is, it is assumed that the family of identities
\begin{equation}
\exp[-\beta' H'_{\Lambda'}(\sigma'|\omega')] \;\bydef\;
\ssum_{\sigma_\Lambda\omega\, \in\, {\cal
C}(\sigma'_{\Lambda'}\omega')}
\exp[-\beta H_\Lambda(\sigma|\omega)] \;,
\label{ft.80}
\end{equation}
gives rise to Hamiltonians $H'$ that can be written in the form
\reff{ft.45} for a suitable interaction that {\em satisfies the
summability condition}\/ \reff{ft.sum}.  The purpose of this paper is
to show that this assumption can be {\em false}\/ at least at low
enough temperatures.  This is the phenomenon of non-Gibbsianness
referred to in the title of this section.  

This lack of Gibbsianness will be proven in the next section by
showing that the renormalized weights fail to satisfy the
quasilocality property \reff{ft.65}, or, more precisely, its
generalized version for spins with infinitely many values
\cite{nondiscrete}.  We shall show that there exists 
one configuration $\omega'$ for which the Boltzmann-Gibbs weight
$\rho(\sigma'|\omega')$ violate the quasilocality condition for some
volume $\Lambda'$ (formed by only two sites!) and some configuration $\sigma'$.
%
%
We emphasize that the existence of {\em one}\/ such a configuration
$\omega'$ is enough to disprove the existence of an acceptable (i.e.\ 
summable) renormalized Hamiltonian.  The summability property
\reff{ft.sum} implies the quasilocality \reff{ft.65} (or its
generalization for spins with infinitely many vaules) for {\em all}
$\omega'$.  Nevertheless, it is natural to wonder how relevant the
phenomenon is from the physical point of view, specially if the
violation of quasilocality involves very atypical
configurations $\omega'$.  We shall comment on this point in Section
\ref{s.non}.

The lack of quasilocality can be interpreted as exhibiting some sort
of ``action at a distance'': Infinitely far away spin flips produce a
sizeable change close to the origin, {\em even when the intermediate
  renormalized spins are frozen in the configuration}\/ $\omega'$.
This is in contrast with the usual behavior in equilibrium statistical
mechanical (Gibbsian behavior) where changes at infinity can propagate
only through fluctuations of intermediate spins.  It is not hard to
imagine the explanation: The fixing of a renormalized configuration
still leaves some fluctuation possible in the system of original spins
compatible with it.  These fluctuations act as ``hidden degrees of
freedom'' that in some instances can bring information from infinity.
This happens when the constrained system of original spins develops
long-range correlations, i.e.\ when it undergoes a {\em phase
  transition}\/.  The argument of next section consists, precisely, in
showing conditions and discussing examples under which such phase
transitions do take place.

\section{Non-Gibbsianness due to momentum transformations}
\label{s.non}

We consider the nearest-neighbor ferromagnetic Ising model in $\scrl=\zed^2$
\begin{equation}
H \;=\; - \sum_{\langle \vv x, \vv y \rangle} \sigma_{\vv x}\,
\sigma_{\vv y} + h \sum_{\vv y} \sigma_{\vv y} \;,
\label{1}
\end{equation}
at low temperatures, that is large $\beta$.
It has been shown that the low-temperature states for this model
under a (local) block-average transformation with even block-sizes are mapped 
onto non-Gibbsian states \cite[Theorem 4.6]{vEFS_JSP}.  A very simple example
of this phenomenon for 1 by 2 blocks was presented in \cite{entbud}.
We shall now prove a similar result for a momentum transformation of the type 
introduced above.  

Let us first sketch some intuition behind our argument.  
In real space, a momentum
transformation looks approximately like an average:
\begin{equation}
\sigma'_{\vv x'} \;=\; \sum_{\vv y\in\zed^d} f(L\vv x'-\vv y)\, 
\sigma_{\vv y}\;,
\label{ft.100}
\end{equation}
where $f$ is the Fourier (anti)transform \reff{ft.ft} of the cutoff
function $\ft f(\vv k)$.  Expressions like \reff{ft.100} have been
studied for example in \cite{gawkup80,gawkup86} (see also
\cite[Appendix 2]{ma76} for a stochastic version).
Even when \reff{ft.100} involves a sum over
all spins $\vv y$ of the lattice, one would expect that 
each $\sigma'_{\vv x'}$ is essentially determined only by
the spins $\sigma_{\vv y}$ inside the block of side $L$ centered at
$L\vv x'$.    Therefore, the mechanism causing non-Gibbsianness for
average transformations \cite[Section 4.3.5]{vEFS_JSP} should apply to
the present case with minor adaptations. 

We will take for our example the identity in one direction and in the other 
direction the soft cutoff function: 
\begin{equation}
\ft f(k) \;=\; \left\{\begin{array}{ll}
cos^{2}(k) & \hbox{for} |k| \leq {\pi \over 2},\\
0 & \hbox{otherwise}\;.
  \end{array}
\right.
\end{equation}

This function integrates out half of the  momenta degrees of freedom
in this direction, which corresponds to taking a (not strictly local) average  
over 
blocks of size $1$ by $2$, centered at sites with even coordinates in the direction in which we renormalize.  

Its Fourier transform is easily computable.
Indeed, $f(0)= {1 \over 4}$, 
$f(2)=f(-2)= {1 \over 8}$,
and 
for all other $n$
\begin{equation}
f(n) = - {2 \over \pi}\, sin\Bigl(n{\pi\over 2}\Bigr) \times {1 \over
  {(n-2)\,n\,(n+2)}} \;.
\end{equation}
(In particular $f(n)=0$ for all even $n\neq 0, \pm 2$).

The initial (and crucial) part of the argument consists in exhibiting
a transformed configuration $\omega'$ such that the corresponding
constrained system of original spins has a phase transition \emph{at
  zero temperature}.  The configuration in question is $\omega'_{\vv
  x}=0$ for all $\vv x\in \zed^2$.  The corresponding original
configurations must, therefore, satisfy the constraint
\begin{equation}
\sum_{l} \  f(l)\, \omega_{2n+l} = 0
\label{gr.1}
\end{equation}
for each $n$ in the direction under consideration.  
We claim that the only 
four groundstates are the 4-periodic configuration (strip state)
\begin{equation}
++--++--++
\end{equation}
and its translates over distances $1$,$2$ or $3$ (while in the other
direction they are of course translation invariant).  It is immediate
to check that these configurations are compatible with the constraint.
Moreover, it is not difficult to check that under the constraint
\reff{gr.1} they are groundstates.

Indeed, assume that we would have a row of 3 identical spins, say
plus, next to each other, then we claim that the above constraint
could not be satisfied. If the middle plus would be on an even site,
say at zero, this would require that
\begin{equation}
\sum_{|l| \geq 3} f(l) \;\geq\; 2\, f(1) \quad \Bigl(= {4 \over {3 \times \pi}}\Bigr)
\end{equation}  
which a simple calculation shows to be impossible.
If, on the other hand, the middle plus would be on an odd site, the interval of
3 plus spins would need to have a minus spin both to its left and to its right
(because of the argument above), and the constraint centered at either the left
or the right site could only be fulfilled if  
\begin{equation}
\sum_{ |l| \geq 3} f(l) \;\geq\; f(0) \quad (= {1 \over 4})
\end{equation}
which again is impossible.  This shows that the constrained system has
multiple, namely four, ground states.

The remaining part of the argument follows closely the presentation in
\cite[Section 4.3.5]{vEFS_JSP}.  There are three additional steps:
\medskip

\noindent
\emph{(1) Existence of a phase transition at nonzero temperature for
  the constrained system.} This follows from a well-known theory
(Pirogov-Sinai theory \cite[Chapter 2]{sin82}, \cite[Appendix
B]{vEFS_JSP}; note that as remarked in \cite{holsla78}, the theory
also applies to systems with constraints.).  There is one extremal
phase associated to each of the four ground states.
\medskip

\noindent
\emph{(2) Selection of the phases of the constrained system via
  block-spin boundary conditions.} This can be done by choosing a
profile $\eta'$ such that if it is imposed in a sufficiently large
(but finite) volume, the configuration deep inside this volume has to
be close to the prescribed ground state.  This is straightforward,
though a little cumbersome to write.  The idea is as follows: Pick
first the ground-state of the constrained system that corresponds to
the phase one wants to select.  Then pick a (very) large volume
$\Delta$, and outside it set the original spins equal to all ``$+$''
or all''$-$''.  Now compute the resulting block-spin configuration
according to \reff{ft.100}.  It will have the property that, for
regions $U$ sufficiently inside $\Delta$, the block spins will be very
close to zero.  The profile we talked about corresponds, then, to the
block spins in some ring around $\Delta$ and in $\Delta\setminus U$.
One sees that by playing with the parity of the boundary of $\Delta$
and changing ``$+$'' to ``$-$'' outside $\Delta$, one can select
alternatively the four phases of the constrained system
\cite{difficult}.
\medskip

\emph{(3) ``Unfixing'' of the spins close to the origin.} This is an
uncomplicated and inessential step.  See the discussion in
\cite[Section 4.2]{vEFS_JSP}.)
\medskip

We see that the argument is insensitive to
the presence of a magnetic field (because the constrained system is
asked to have small magnetization), 
thus we are proving non-Gibbsianness for low temperatures but {\em
arbitrary magnetic field}\/.

Moreover, we remark that in case the blocks are large, having small
magnetisation in a block at low temperatures represents a large
fluctuation from the typical behavior, in which the magnetisation is
either positive or negative of order $O(L^d)$. Renormalized effective
interactions are known not to be adequate to describe such large
values of the fluctuation field \cite{gawkup86}.  This large
(fluctuation) field region can in good cases be treated by polymer
expansions.

\section{Comments and conclusions}

The present example of non-Gibbsianness as a consequence of
momentum-space transformations confirms the suspicion of Griffiths
\cite{gri81} that "no peculiarities of this sort have been
found\ldots, which may merely reflect the fact that no one has looked
for them!".  Nevertheless, one should not draw too radical conclusions
from this occurrence.  On the practical side, the main implication of
non-Gibbsianness is that one has to be very careful in designing
renormalization group transformations.  This is in complete agreement with
what the founders and various practitioners of Renormalization-Group
methods have been saying all along.

Indeed, already Wilson and Kogut in their classic review
emphasized 

\begin{quotation}
Otherwise, [that is 
non-perturbatively], the locality of [the renormalized interactions] is a 
non-trivial problem, which will not be discussed further
\cite[p.~145]{Wilson-Kogut}. 
\end{quotation}

And more explicitly, M.~E.~Fisher in his ``Renormalization Group
Desiderata'', listed the conditions needed for a succesful
renormalization scheme in Hamiltonian space:

\begin{quotation}
A Renormalization Group for a space of Hamiltonians should satisfy the
following:  A) Existence {\em in} the thermodynamic limit,\ldots
C) Spatial locality \ldots, one should be able to identify the same
regions of space and associated local variables before and after the
transformation \cite[Section 5.4.2]{fis83}
\end{quotation}

Our example adds to the numerous instances showing that perversely or
sloppily designed transformations can lead people into trouble.  As
N.~Goldenfeld points out in his book {\em Lectures on Phase
  Transitions and the Renormalization Group}\/ \cite[p.~268]{gol92}:
``It is dangerous to proceed without thinking about the physics''.
The moral is, then, that renormalization transformations must be
carefully crafted and case-tailored.  Already Wilson, as quoted in
\cite[p.~ 492]{nievan76}, warned: ``One can not write a renormalization
cookbook''.

On the foundational side, examples like the present one confirm the
view expressed by Benfatto and Gallavotti \cite{bengal95} in the
opening sentence of their book, {\em Renormalisation Group}\/: ``The
notion of Renormalisation Group is not well-defined''.  It is clear
that the mathematical formalization of the method requires much more
than a naive approach in terms of Hamiltonians and flows of coupling
constants.  We mention two directions of work which can potentially
lead to a better mathematical understanding of the renormalization
group framework. 

On the one hand, the connection with the large-field problem suggests
to combine renormalization group ideas with geometrical expansions
---cluster or polymer expansions--- to circumvent the ill-definedness
of the renormalized Hamiltonian.  These expansions have indeed been
succesfully applied in the rigorous control of renormalization group
transformations of unbounded-spin systems \cite[and references
therein]{gawkup80,gawkup86}.  A related approach, for bounded-spin
systems, resorted to the renormalization of Peierls-like contours
\cite{gawkotkup87}.  

On the other hand, the more recent program started by Dobrushin
\cite{dob95,dobshl97,lormae97,brikuplef98,dobshl98} studies
non-Gibbsianness with techniques borrowed from the treatment of
Griffiths singularities.  He proposes a more general class of allowable Hamiltonians, leading to the notion of ``weak
Gibbsianness'' as the right framework for a unified treatment.

We think our result illustrates and clarifies to some extent the
reason why finding a good renormalization group scheme is such a non-trivial
task, not only for strictly local but also for only approximately
local transformations.  We produced an example in the low-temperature
regime, but the fact that the mechanisms of non-Gibbsianness are so
similar for real-space and momentum-space transformations, leads us to
the conjecture that, as in real space, also in momentum-space one
cannot trust that in general the critical region is free of problems.

\section*{Acknowledgments}
We thank various colleagues, and in particular Michael Fisher, for
pointing out to us that our earlier results applied to position-space
transformations only,
and that the question whether momentum-space transformations 
behaved in a similar manner deserved consideration. Our
treatment owes much to our earlier collaboration with Alan Sokal. 
R.F. thanks the Rijksuniversiteit Groningen for hospitality while this
work was performed, and the FOM-SWON samenwerkingsverband 
Mathematische Fysica for supporting his visit.
The work of R.F. was partially supported by FAPESP (grant 95/0790-1,
Projeto Tem\'atico ``Fen\^omenos cr\'\i ticos e processos evolutivos e
sistemas em equil\'\i brio''), CNPq (grant 301625/95-6) and FINEP
(N\'ucleo de Excel\^encia ``Fen\'omenos cr\'\i ticos em probabilidade
e processos estoc\'asticos'', PRONEX-177/96), and the work of
A.C.D.v.E. by EU contract CHRX-CT93-0411.


\begin{thebibliography}{10}
\bibitem{nondiscrete}
Formally, quasilocality for spins with infinitely many values is defined as
  follows. Let us denote ${\cal O}_{\delta\,,\,\Lambda}(\sigma)$ the set of
  configurations differing less than $\delta$ at sites in $\Lambda$: $$ {\cal
  O}_{\delta\,,\,\Lambda}(\sigma) \;=\; \Bigl\{\widetilde\sigma \,:\,
  \left|\widetilde\sigma_{\vv x} - \sigma_{\vv x}\right| \le \delta \;,\; \vv
  x\in \Lambda \Bigr\} \;. $$ Then $H_\Lambda(\sigma|\,\cdot\,)$ is quasilocal
  at $\omega$ if for any $\epsilon>0$ and ${\cal
  O}_{\delta\,,\,\Lambda}(\sigma)$ there exists a $\lambda>0$ and
  $\Gamma\subset\zed^d$ such that $$ \left| H_\Lambda(\widetilde\sigma |
  \widetilde \omega) - H_\Lambda(\sigma|\omega)\right| \;<\; \epsilon $$ for
  every $\widetilde\sigma \in {\cal O}_{\delta\,,\,\Lambda}(\sigma)$ and
  $\widetilde\omega \in {\cal O}_{\lambda\,,\,\Gamma}(\omega)$.
\end{thebibliography}

\end{document}